\documentclass{PoS}
\usepackage{cite}
\usepackage{amsmath,braket}

\title{Long-distance contributions to flavour-changing processes}

\ShortTitle{Long-distance effects}

\author{\speaker{C.T.Sachrajda}\\ 
        School of Physics and Astronomy, University of Southampton, Southampton SO17 1BJ, UK\\
        E-mail: \email{cts@soton.ac.uk}}

\abstract{Standard lattice calculations in flavour physics or in studies of hadronic structure are based on the evaluation of matrix elements of local composite operators between hadronic states or the vacuum. In this talk I discuss developments aimed at the computation of long-distance, and hence non-local, contributions to such processes. 
In particular, I consider the calculation of the $K_L$-$K_S$ mass difference $\Delta m_K=m_{K_L}-m_{K_S}$ and the amplitude for the rare-kaon decay processes $K\to\pi\ell^+\ell^-$, where the lepton $\ell=e$ or $\mu$. Lattice calculations of  the long-distance contributions to the indirect $CP$-violating parameter $\epsilon_K$ and to the rare decays $K\to\pi\nu\bar\nu$ are also beginning.
Finally I discuss the possibility of including $O(\alpha)$ electromagnetic effects in computations of leptonic and semileptonic decay widths, where the novel feature is the presence of infrared divergences. This implies that contributions to the width from processes with a real photon in the final state must be combined with those with a virtual photon in the amplitude so that the infrared divergences cancel by the Bloch-Nordsieck mechanism. I present a proposed procedure for lattice computations of the $O(\alpha)$ contributions with control of the cancellation of the infrared divergences.} 

\FullConference{The 32nd International Symposium on Lattice Field Theory,\\
		23-28 June, 2014\\
		Columbia University New York, NY}

\begin{document}

\section{Introduction}\label{sec:intro}
Our satisfaction at the discovery of the Higgs Boson is tempered, only temporarily we hope, by the absence of the discovery of new physics at the Large Hadron Collider (LHC). A key complementary tool to large $E_T$ searches for signs of physics Beyond the Standard Model (BSM) is precision flavour physics. If after more detailed analysis of LHC data generated before the current shutdown, or of LCH14 data generated in the future, new particles are discovered then precision flavour physics will be necessary to unravel the underlying theoretical framework. We should also not underestimate the discovery potential of flavour physics; an unambiguous inconsistency between the standard model prediction of a weak-decay rate or CP-asymmetry and the corresponding experimental measurement would signal new BSM particles and couplings. Indeed it is perhaps surprising that no such unambiguous inconsistencies have arisen up to now. Precision flavour physics requires quantitative control of hadronic effects for which lattice QCD simulations are essential.

At this conference we have seen the continued, hugely impressive, improvement in the precision of lattice calculations for a wide range of quantities. As an example of the improved precision consider the renormalisation group invariant $B_K$ parameter of neutral kaon mixing. At the 2013 conference on high-energy physics~\cite{Sachrajda:2014hfa} I quoted $\hat B_K=0.766(10)$ from the compilation of the Flavour Physics Lattice Averaging Group~(FLAG)\,\cite{Aoki:2013ldr}. Twenty years earlier at the same conference I quoted $\hat B_K=0.8(2)$~\cite{Sachrajda:1993dd}.
Not only has the error decreased by a factor of 20, but our confidence in estimating the uncertainty is much greater that it could ever be in the era of quenched simulations with pion masses in the region of 500\,MeV - 1\,GeV. For further examples of impressive improvement in precision see table\,\ref{tab:decayconstants} below for the FLAG compilation of pseudoscalar leptonic decay constants. The twofold challenge now for our community is to maintain the improvement in precision in the calculation of standard quantities and to increase the range of physical quantities which can be made accessible to lattice simulations. With this in mind, in this talk I will discuss the extension of the evaluation of matrix elements to include non-local effects. In standard computations we evaluate matrix elements of the form $\bra{0} O(0)\ket{h}$ or $\bra{h_2} O(0)\ket{h_1}$ where $O$ is a local composite operator and $h,h_{1,2}$ are hadronic states. For the quantities discussed in this talk, the matrix elements are of time-ordered products of two local composite operators integrated over their positions (see eq.(\ref{eq:genericme})).

I will discuss 3 topics in which the contributing matrix elements are of non-local operators involving long-distance effects:
\vspace{-0.06in}\begin{enumerate}
\item The calculation of the $K_L$-$K_S$ mass difference, $\Delta m_K=m_{K_L}-m_{K_S}$. This work is being performed with colleagues from the RBC-UKQCD collaboration and the current status was reviewed by Z.\,Bai at this conference~\cite{Bai:2014hta}.
\vspace{-0.1in}\item The study of rare kaon decays $K\to\pi\ell^+\ell^-$ or $K\to\pi\nu\bar{\nu}$. This work is also being performed with colleagues from the RBC-UKQCD collaboration and has been summarised at this conference by X.\,Feng~\cite{xu}.
\vspace{-0.1in}\item The evaluation of the electromagnetic corrections to weak decay amplitudes~\cite{Carrasco:2015xwa}. 
\end{enumerate}
Before discussing these specific processes however, I start by considering a common issue for all such studies of long-distance effects, i.e. how to construct the initial and final states and yet be able to integrate over the positions at which the operators are inserted.

\subsection{The fiducial volume}

\begin{figure}[t]
\begin{center}
\includegraphics[width=0.8\hsize]{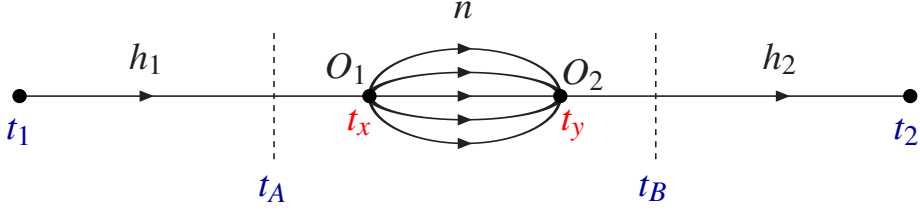}
\end{center}
\caption{A schematic diagram to illustrate the integrated matrix element in eq.\,(\protect \ref{eq:genericme}). The hadrons $h_1$ and $h_2$ are created and annihilated by interpolating operators located at $t_1$ and $t_2$ respectively and the time positions of the operators $O_{1,2}$ are integrated between $t_A$ and $t_B$. $n$ represents a generic intermediate state between the two operators.\label{fig:fiducial}}
\end{figure}

The generic non-local quantity which we will be considering when studying $\Delta m_K$ or rare kaon decays is of the form
\begin{equation}\label{eq:genericme}
\int d^4x\, \int d^4 y\ \langle\, h_2\,|\,T\{O_1(x)\,O_2(y)\}\,|\,h_1\rangle\,,\end{equation}
where $O_{1,2}$ are composite operators and $|\,h_{1,2}\rangle$ are single-hadron states\footnote{One can also exploit translational invariance and set either $x$ or $y$ to be the origin for example.}. In infinite spatial and temporal Minkowski volumes, one first constructs the asymptotic in- and out-states $\ket{h_1}$ and $\bra{h_2}$ and then integrates over all $x$ and $y$, and in particular over the times $t_{x,y}$. A corresponding practical procedure in finite-volume Euclidean simulations is to integrate over a large subinterval in time $t_A\le t_{x,y}\le t_B$, but to create $h_1$ and to annihilate $h_2$ well outside of this region to allow for the isolation of the single-particle external state. We call the 4-dimensional volume between the time slices at $t_A$ and $t_B$, the \textit{fiducial volume}. This procedure will be discussed in some detail in sections\,\ref{sec:DeltamK} and \ref{sec:rarekaondecays} below.

\section{The $K_L$-$K_S$ mass difference}\label{sec:DeltamK}
The material in this section is based on the work of the RBC and UKQCD collaborations reported in~\cite{Christ:2012se,Bai:2014cva} and updated in Z.Bai's talk at this conference~\cite{Bai:2014hta}. The $K_L$-$K_S$ mass difference is known with excellent precision,  
\begin{equation}\label{eq:DeltamKexperimental}
\Delta m_K\equiv m_{K_L}- m_{K_S}=3.483(6)\times 10^{-12}~\mathrm{MeV}
\end{equation}
and it is a challenge for our community to compute the non-perturbative QCD effects in order to be able either to reproduce such a value in the standard model or to convincingly demonstrate an inconsistency which would signal the presence of new physics. Such a tiny value of $\Delta m_K$ can potentially set strong constraints on physics beyond the standard model and historically led to the suggestion of the existence of the charm quark with an estimate of the corresponding mass scale~\cite{Mohapatra:1968zz,Glashow:1970gm,Gaillard:1974hs}.

Within the standard model, $\Delta m_K$ arises from $K^0$--$\bar{K}^0$ mixing at second order in the weak interactions: 
\begin{equation}\label{eq:DeltamKtheory}
\Delta M_K=2{\cal P}~\sum_\alpha\,\frac{\langle\,\bar{K}^{\,0}\,|\,H_W\,|\alpha\rangle\,\langle\alpha\,|H_W\,|\,K^0\rangle}
{m_K-E_\alpha}\,,\end{equation}
where the sum over intermediate states $|\alpha\rangle$ includes an energy-momentum integral and ${\cal P}$ indicates that the principal value of the pole in (\ref{eq:DeltamKtheory}) is to be taken.

For $\Delta m_K$, in figure\,\ref{fig:fiducial} and eq.\,(\ref{eq:genericme}) we take $h_1=K^0$, $h_2=\bar{K}^0$ and both $O_{1,2}$ to be the effective $\Delta S=1$ weak Hamiltonian $H_W$.  Its lattice determination begins with the evaluation of the integrated correlation function:
\begin{equation}\label{eq:c4def}
C_4(t_A,t_B;t_1,t_2)=\frac12\sum_{t_y=t_A}^{t_B}\sum_{t_x=t_A}^{t_B}\langle 0\,|\,T\left\{\bar{K}^0(t_2)\,H_W(t_y)\,H_W(t_x)\,\bar{K}^0(t_1)
\right\}|\,0\rangle\,,
\end{equation}
where the integral over the 3-dimensional spatial positions of all 4 operators is implicit. Inserting complete sets of states between the operators and integrating $t_{x,y}$ between $t_A$ and $t_B$ we find:
\begin{eqnarray}
C_4(t_A,t_B;t_1,t_2)&=&|Z_K|^2e^{-m_K(t_2-t_1)}\sum_n\,\frac{\langle\bar{K}^0\,|\,{\mathcal H_W}\,|\,n\rangle\,\langle n\,|\,{\mathcal H_W}\,|\,K^0\rangle}{(m_K-E_n)^2}\,\times\nonumber
\\ &&\hspace{0.75in}
\bigg\{e^{(M_K-E_n)T}-(m_K-E_n)T-1\bigg\}\,.\label{eq:c4Tdep}\end{eqnarray}
where $T=t_B-t_A+1$. From the coefficient of $T$ we can therefore obtain
\begin{equation} 
\Delta m_K^{\mathrm{FV}}\equiv2 \sum_n\,\frac{\langle\bar{K}^0\,|\,H_W\,|\,n\rangle\,\langle n\,|\,H_W\,|\,K^0\rangle}{(m_K-E_n)}\,,
\label{eq:DeltamKFV}\end{equation}
where the superscript $\scriptsize{\textrm{FV}}$ reminds us that the quantity has been evaluated on a finite volume. We will discuss the finite-volume corrections below.

A generic feature in the evaluation of matrix elements of bilocal operators is the presence of terms which (potentially) grow exponentially in $T$. In this case we see from eq.\,(\ref{eq:c4Tdep}) that this will be the case 
if there are intermediate states $n$ with energies $E_n<m_K$.  Among such terms are the $\pi^0$ and vacuum intermediate states whose contribution can be identified and subtracted numerically (leading to some loss of precision). 
Alternatively we can exploit the fact that one can add terms proportional to the scalar density $\bar{s}d$ or the pseudoscalar density $\bar{s}\gamma^5d$ without changing physical quantities, $\Delta m_K$ in particular, and choose the corresponding coefficients so that $\langle \pi^0\,|\,H_W\,|K\rangle$ and $\langle 0\,|\,H_W\,|K\rangle$ are both zero, where $H_W$ is now the Hamiltonian after the subtraction. In practice it may not always be optimal to chose the coefficients of the densities in this way, e.g. in \cite{Bai:2014hta} for simulations at unphysical masses it was more effective to remove the contribution of the $\eta$-meson instead of the pion. In addition to the vacuum and single pion states, for physical quark masses there are also two-pion (and three-pion) contributions with energies $E_{\pi\pi}<m_K$ whose exponentially growing contributions have to be subtracted numerically. Moreover the number of states with energies less than $m_K$ grows as the volume increases, although in the foreseeable future it is likely that there will only be one or two such states. 

The $\Delta S=1$ effective Weak Hamiltonian takes the form:
\[
H_W=\frac{G_F}{\sqrt{2}}\sum_{q,q^{\prime}=u,c}V_{qd}V^{*}_{q^{\prime}s}(C_1Q_1^{qq^{\prime}}+C_2Q_2^{qq^{\prime}})\]
where  the $\{Q_i^{qq{\prime}}\}_{i=1,2}$ are current-current operators, defined as:
\begin{equation}
Q_1^{qq{\prime}}=(\bar{s}_i\gamma^\mu(1-\gamma^5)d_i)~(\bar{q}_j\gamma^\mu(1-\gamma^5)q^{\prime}_j)\quad\textrm{and}\quad
Q_2^{qq{\prime}}=(\bar{s}_i\gamma^\mu(1-\gamma^5)d_j)~(\bar{q}_j\gamma^\mu(1-\gamma^5)q^{\prime}_i)\,.\label{eq:Q1Q2}
\end{equation}
$i$ and $j$ are colour labels and the spinor indices are contracted within each set of brackets.

\begin{figure}[t]
\begin{center}
\includegraphics[width=0.5\hsize]{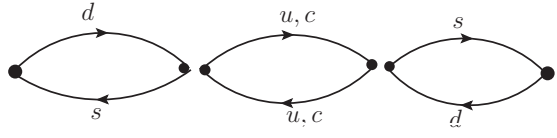}
\end{center}
\caption{Contribution to the correlation function $C_4$.\label{fig:c41}}
\end{figure}

The ultraviolet component of the evaluation of $\Delta m_K$ is particularly benign, provided that one performs the computation in the four-flavour theory and includes the charm quark. Of course each of the two weak Hamiltonians has to be renormalised and I assume that this has been done. In addition, one might expect that additional divergences will arise from regions of the integration space when $x$ and $y$ approach each other. For example in the inner loop of the diagram in Fig.\,\ref{fig:c41}, the $u$-quark contribution can be seen by power counting to be quadratically divergent. Not only is the quadratic divergence cancelled by the GIM mechanism when the charm quark is included but, because of the chiral structure of the operators $Q_{1,2}$ in eq.\,(\ref{eq:Q1Q2}), so do the logarithmic ones~\cite{Christ:2012se}. The short distance contributions come from distances of $O(1/m_c)$ and not of the order of the ultraviolet cutoff, the lattice spacing  $a$.

\begin{figure}[t]
\begin{center}
\includegraphics[width=0.4\hsize]{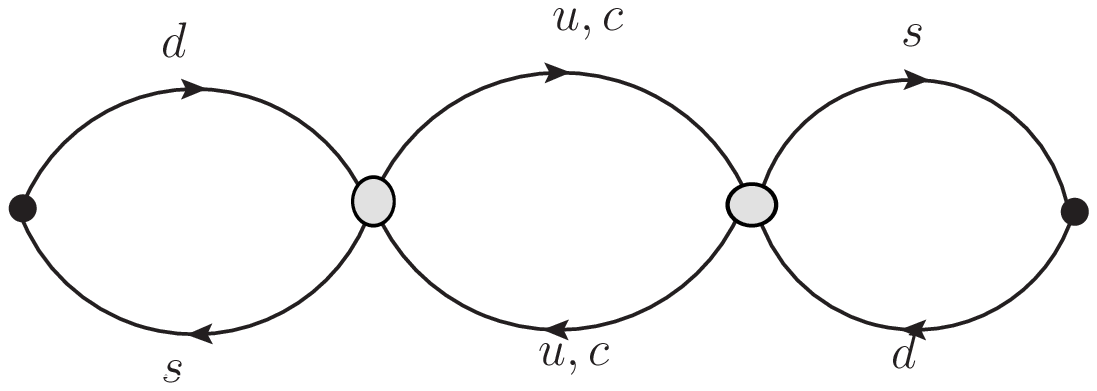}\quad
\includegraphics[width=0.35\hsize]{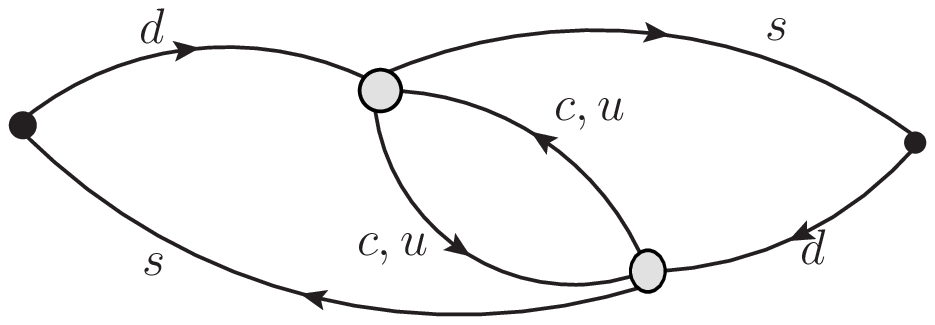}\end{center}
\vspace{-0.2in}\hspace{1.7in}Type 1\hspace{1.8in}Type 2\\ 
\begin{center}
\vspace{-.3in}\includegraphics[width=0.4\hsize]{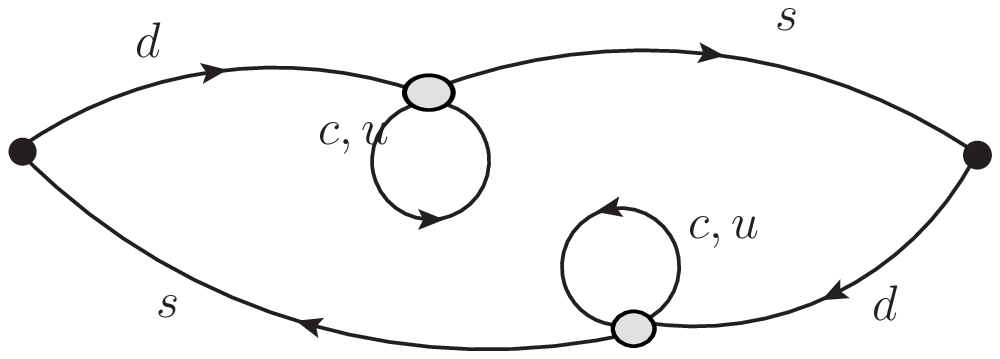}\quad
\includegraphics[width=0.4\hsize]{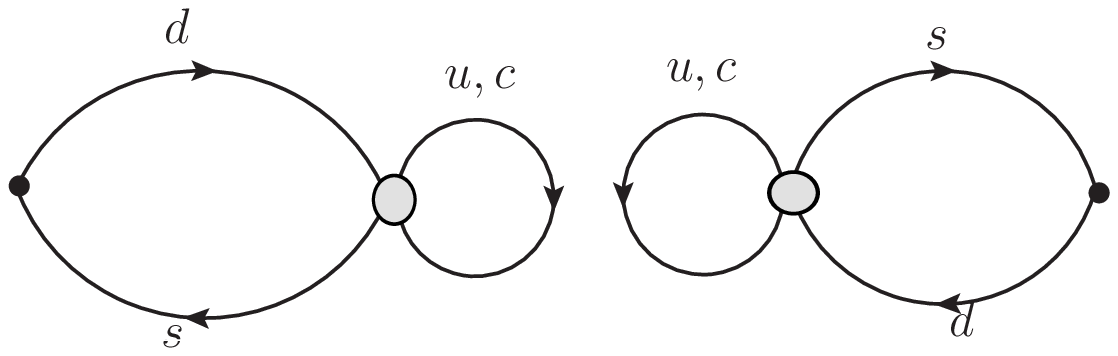}\quad
\end{center}
\vspace{-0.2in}\hspace{1.7in}Type 3\hspace{2in}Type 4
\caption{The four types of diagram contributing to the correlation function $C_4$ from which $\Delta m_K$ is obtained.\label{fig:DeltamKtypes}}
\end{figure}

The RBC-UKQCD collaboration has been performing exploratory calculations of $\Delta m_K$, starting with a calculation of only the diagrams of Types 1 and 2 on a $16^3$ lattice with $a^{-1}=1.73\,$GeV and $m_\pi\simeq420\,$MeV~\cite{Christ:2012se}. In a more recent study we have performed a full calculation of all graphs, on a $24^3$ lattice with the same lattice spacing, with the domain wall fermion (DWF) and the Iwasaki gauge action, and with $m_\pi=330\,$MeV, $m_K=575$\,MeV, $m_c^{\overline{\mathrm{MS}}}(2\,\mathrm{GeV})=949$\,MeV and $am_\mathrm{res}=0.00308(4)$~\cite{Bai:2014cva}. At these unphysical quark masses we find
\begin{equation}
\Delta m_K=3.19(41)(96)\times 10^{-12}\,\mathrm{MeV}\,,\label{eq:DeltamKPRL}\end{equation}
to be compared to the physical value of 3.483(6)$\times 10^{-12}$\,MeV.
The agreement with the physical value may well be fortuitous, but it is nevertheless reassuring to obtain results of the correct order. The systematic error is dominated by discretisation effects related to the charm quark mass, which we estimate at 30\%. In this computation $m_K<2m_\pi$ and so we do not have any exponentially growing contributions from two-pion intermediate states.

\begin{table}[t]
\begin{center}
\begin{tabular}{|c|c|c|c|c|c|}\hline
$m_\pi$&$m_K$&$m_c$&$a^{-1}$&$L$&no. of configs.\\ 
\rule[-\baselineskip]{0pt}{12pt}
171\, MeV&492\,MeV & 592/750\,MeV & 1.37\,GeV&4.6\,fm &212\\ \hline
\end{tabular}\end{center}
\vspace{-.1in}\caption{Parameters of the simulation used in~\cite{Bai:2014hta} to calculate $\Delta m_K$.\label{tab:ziyuan}}
\end{table}

At this conference, Z.Bai presented a stays report of work in progress by the RBC-UKQCD collaboration~\cite{Bai:2014hta}. This is a calculation on the $32^3\times 64$ coarse lattice ($a^{-1}$ = 1.37(1)\,GeV) with domain wall fermions and the DSDR gauge action (see \cite{Arthur:2012opa} for details of the ensemble) and which had been used in the first computation of $K\to(\pi\pi)_{I=2}$ decay amplitudes~\cite{Blum:2011ng,Blum:2012uk}, where the subscript ${\scriptsize I=2}$ indicates that the total isospin of the two-pion final state is 2. The parameters of the simulation are presented in table~\ref{tab:ziyuan}. The new feature of this study is that now $m_K>2m_\pi$ which allows us to study the effect of the two-pion intermediate state. The preliminary results of this calculation were $\Delta m_K=(4.6\pm 1.3)\times 10^{-12}$\,MeV when $m_c=750\,\textrm{MeV}$ and $\Delta m_K=(3.8\pm 1.7)\times 10^{-12}$\,MeV when $m_c=592\,\textrm{MeV}$. Only statistical errors are shown. In this calculation it is seen that the contributions from the $\pi\pi$ intermediate state are very small (3-4\%). 

From this series of investigations we learn that a calculation of $\Delta m_K$ at physical kinematics and on ensembles with unquenched charm quarks will be possible in the very near future. For the prospects for the calculation of the long-distance contributions to the indirect CP-violating parameter $\varepsilon_K$ see~\cite{Christ:2014qwa}.

I end this section with a brief discussion of the finite-volume corrections which relate $\Delta m_K^\textrm{FV}$ in eq.\,(\ref{eq:DeltamKFV}), extracted from the correlation function $C_4$ (eqs.\,(\ref{eq:c4def}) and (\ref{eq:c4Tdep})), to the physical value of $\Delta m_K$.  Because of the pole at $E_n=m_K$ in eq.\,(\ref{eq:DeltamKFV}) this requires an extension of the Lellouch-L\"uscher formula for $K\to\pi\pi$ decays~\cite{Lellouch:2000pv}. Assuming that the dominant contribution comes from $s$-wave rescattering of the two pions the relation is~\cite{Christ:2014qaa,CFMS}
\begin{eqnarray}\Delta m_K&=&\Delta m_K^\mathrm{FV}-2\pi\,\mbox{}_V\hspace{-1pt}\langle\bar{K}^0\,|\,H_W\,|\,n_0\rangle_V\,
\mbox{}_V\!\langle n_0\,|\,H_W\,|\,K^0\rangle_V\,\left[\cot \pi h\,\frac{dh}{dE}\right]_{m_K}\label{eq:CMS1}\\ 
&=&\Delta m_K^\mathrm{FV}-2\pi\,\langle\bar{K}^0\,|\,H_W\,|\,n_0\rangle\,
\!\langle n_0\,|\,H_W\,|\,K^0\rangle\,\cot [\pi h(m_K,L)]\,,
\label{eq:CMS2}\end{eqnarray} 
where $h(E,L)\pi\equiv\phi(q)+\delta_0(k)$, $\delta_0$ is the $s$-wave $I=0$ $\pi\pi$ phase shift~\footnote{Because of the $\Delta I=1/2$ rule, we neglect the contribution of the isospin $2$ intermediate state, but this can be simply included by adding the corresponding term on the right-hand side of eqs.\,(\ref{eq:CMS1}) and (\ref{eq:CMS2}).} and $\phi$ is a kinematic function so that L\"usher's quantisation condition is $h(E_n,L)=n$~\cite{Luscher:1990ux}. $|n_0\rangle$ is a two-pion state with energy $E_{n_0}=m_K$ and the subscript ${\scriptsize V}$ in eq.\,(\ref{eq:CMS1}) denotes that the matrix elements are those in the finite-volume. More precisely
\begin{equation}
\left(\frac{dh}{dE}\right)_{\hspace{-0.02in}m_K}\,\mbox{}_V\hspace{-1pt}\langle\bar{K}^0\,|\,H_W\,|\,n_0\rangle_V\,
\mbox{}_V\!\langle n_0\,|\,H_W\,|\,K^0\rangle_V=\langle\bar{K}^0\,|\,H_W\,|\,n_0\rangle\,
\!\langle n_0\,|\,H_W\,|\,K^0\rangle\,,
\end{equation}
where on the right-hand side we have the infinite-volume matrix elements. The kaon state $|K\rangle$ has been normalised relativistically throughout this discussion so that $dh/dE$ is simply the two-pion contribution to the Lellouch-L\"uscher factor.

\section{Rare Kaon Decays}\label{sec:rarekaondecays}

Rare kaon decays which are dominated by short-distance flavour-changing neutral current (FCNC) processes, $K\to\pi\nu\bar\nu$ decays in particular, provide a potentially valuable window on new physics at high-energy scales. The decays $K_L\to\pi^0 e^+e^-$ and $K_L\to\pi^0\mu^+\mu^-$ are also considered promising because the long-distance effects are reasonably under control using ChPT~\cite{Mescia:2006jd}. 
They are sensitive to different combinations of short-distance FCNC effects and hence in principle provide additional discrimination to the neutrino modes. A challenge for the lattice community is therefore to calculate the long-distance effects reliably. The existing phenomenology on rare kaon decays is based largely on SU(3)$_\mathrm{L}\times$SU(3)$_\mathrm{R}$ ChPT and lattice calculations will also provide the opportunity for checking the range of validity of ChPT and evaluating the corresponding Low Energy Constants.  

As an example consider the decay $K_L\to\pi^0\ell^+\ell^-$ which has three main contributions to the amplitude~\cite{Mescia:2006jd},
\vspace{-0.1in}\begin{enumerate}
\item[(i)] short distance contributions corresponding to matrix elements of the local operators\\ $(\bar{s}\gamma_\mu d)(\bar\ell\gamma^\mu\ell)$ and $(\bar{s}\gamma_\mu d)(\bar\ell\gamma^\mu\gamma_5\ell)$. The hadronic (non-perturbative QCD) contribution is simply given by the form-factors of semileptonic $K\to\pi\ell\bar{\nu}_\ell$ decays, which are known to much better precision than the remaining contributions.
\item[(ii)] long-distance indirect CP-violating contribution from the CP-even component of $K_L$,\\ $A_{ICPV}(K_L\to\pi^0\ell^0\ell^-)=\epsilon A(K_1\to\pi^0\ell^+\ell^-)$ and 
\item[(iii)] the two-photon CP-conserving contribution $K_L\to\pi^0(\gamma^\ast\gamma^\ast\to\ell^+\ell^-)$. 
 \end{enumerate}

A summary of the corresponding phenomenology is presented in Ref.\,\cite{Cirigliano:2011ny}. For example the branching ratios for the CP-violating (CPV) components are written as:
\begin{eqnarray}
\mathrm{Br}(K_L\to\pi^0 e^+e^-)_{\mathrm{CPV}}&=&10^{-12}\,\times\left\{15.7|a_S|^2\pm 6.2|a_S|\,\left(\frac {\mathrm{Im}\,\lambda_t}{10^{-4}}\right)+2.4\,\left(\frac {\mathrm{Im}\,\lambda_t}{10^{-4}}\right)^2
\right\}\\
\mathrm{Br}(K_L\to\pi^0 \mu^+\mu^-)_{\mathrm{CPV}}&=&10^{-12}\,\times\left\{3.7|a_S|^2\pm 1.6|a_S|\,\left(\frac {\mathrm{Im}\,\lambda_t}{10^{-4}}\right)+1.0\,\left(\frac {\mathrm{Im}\,\lambda_t}{10^{-4}}\right)^2
\right\}\,,
\end{eqnarray}
where $a_S$ is the (unphysical) amplitude for the decay $K_S\simeq K_1\to \pi^0\ell^+\ell^-$ at momentum transfer $q^2=0$. Using ChPT-based phenomenology, $|a_S|=1.06^{+0.26}_{-0.21}$ but the sign of $a_S$ is unknown~\cite{Cirigliano:2011ny}. One goal of future lattice calculations is the determination of $a_S$, together with other similar quantities. In addition however, we will be able to vary the external momenta and study the behaviour of the amplitude as a function of $q^2$. Using partially twisted boundary conditions~\cite{Sachrajda:2004mi,Bedaque:2004ax}, this has been done very successfully for the form factors in $K_{\ell 3}$ decays~\cite{Boyle:2007wg}. See~\cite{Garron:2014pxa} for a review of the current status of the calculations of $K_{\ell 3}$ decay amplitudes.  

From the above discussion we learn that we need to compute the amplitudes for the CP-conserving decays $K_S\to\pi^0\ell^+\ell^-$ and $K^+\to\pi^+\ell^+\ell^-$ and start by considering~\cite{Isidori:2005tv} 
\begin{equation} 
T_i^\mu=\int d^4x\,e^{-iq\cdot x}\,\langle \pi(p)\,|\, \mathrm{T}\{J^\mu_{\mathrm{em}}(x)\,Q_i(0)\,\}\,|\,K(k)\rangle\,,
\end{equation}
where $Q_i$ ($i=1,2$) are operators in the effective Hamiltonian (see (\ref{eq:Q12})) and $J^\mu_{\mathrm{em}}$ is the electromagnetic current. Electromagnetic gauge invariance implies that $T_i^\mu$ takes the form
\begin{equation}\label{eq:Ti}
T_i^\mu=\frac{\omega_i(q^2)}{(4\pi)^2}\,\left\{q^2(p+k)^\mu-(m_K^2-m_\pi^2)\,q^\mu\right\}\,.\end{equation}
It is the form factor $\omega(q^2)$ which will be the output of the calculation. The computation will proceed in a similar way to the evaluation of $\Delta m_K$, by inserting the interpolating operators for the initial kaon and final pion states at times which are sufficiently far from the "fiducial volume", i.e. the range of integration over which the time of the insertion of the current is integrated.  

Although the lattice computation does not rely on ChPT nevertheless, since most of the existing phenomenology is performed in the ChPT framework, it may be useful to compute the necessary low energy constants.
The LECs $a_+$ and $a_S$ are defined by
\begin{equation}\label{eq:a+s}
a=\frac{1}{\sqrt{2}}\,V_{us}^\ast V_{ud}\left\{C_1\omega_1(0)+C_2\omega_2(0)+\frac{2N}{\sin^2\theta_W}\,f_+(0)C_{7V}\right\} 
\end{equation}
where $Q_{1,2}$ are the two current-current GIM subtracted operators and the $C_i$ are the Wilson coefficients, ($C_{7V}$ is the coefficient of ($\bar{s}\gamma_\mu d)(\bar{l}\gamma^\mu l)$)~\cite{D'Ambrosio:1998yj}.
An interesting target for the lattice calculations is to check the validity of the phenomenological values: $a_+=-0.578\pm 0.016$ and $|a_S|=1.06^{+0.26}_{-0.21}$, as well as to determine the sign of $|a_S|$. 

The generic non-local matrix elements which we need to evaluate are (in Minkowski space)
\begin{eqnarray}
X&\equiv&\int_{-\infty}^\infty\,dt_x\,d^3x\ \langle\pi(p)\,|\,\mathrm{T}\left[\,J_\mu(0)\,H_W(x)\,\right]\,|K(\vec{0})\rangle\\ 
&=&i\,\sum_n\,\frac{\langle\pi(p)\,|\,J_\mu(0)\,|n\rangle\,\langle n\,|H_W(0)\,|\,K(\vec{0})\rangle}{m_K-E_n+i\epsilon}
-i\,\sum_{n_s}\,\frac{\langle\pi(p)\,|\,H_W(0)\,|n_s\rangle\,\langle n_s\,|J_\mu(0)\,|\,K(\vec{0})\rangle}{E_{n_s}-E_\pi+i\epsilon}\,.
\end{eqnarray}
$J_\mu$ represents a vector or axial, electromagnetic or weak, current and 
$\{|n\rangle\}$ and $\{|n_s\rangle\}$ represent complete sets of non-strange and $S=1$ strange states.
In Euclidean space we envisage calculating correlation functions of the form
\begin{equation} \int_{-T_a}^{T_b}dt_x\,\langle\phi_\pi(\vec{p},t_\pi)\,\mathrm{T}\left[\,J_\mu(0)\,H_W(t_x)\,\right]\,\phi^\dagger_K(\vec{0},t_K)\,\rangle\equiv \sqrt{Z_K}\,\frac{e^{-m_K|t_K|}}{2m_K}\,X_E\,
\sqrt{Z_\pi}\,\frac{e^{-E_\pi t_\pi}}{2E_\pi}\,,
\label{eq:XE}\end{equation}
where $\phi_\pi$ and $\phi_K$ are interpolating operators for the pion and kaon respectively, $X_E=X_{E_-}+X_{E_+}$ and 
\begin{eqnarray}
X_{E_-}&=&-\,\sum_n\,\frac{\langle\pi(p)\,|\,J_\mu(0)\,|n\rangle\,\langle n\,|H_W(0)\,|\,K\rangle}{m_K-E_n}\left(1-e^{(m_K-E_n)T_a}\right)
\quad\mathrm{and}\label{eq:XE-}\\ 
X_{E_+}&=&\sum_{n_s}\,\frac{\langle\pi(p)\,|\,H_W(0)\,|n_s\rangle\,\langle n_s\,|J_\mu(0)\,|\,K\rangle}{E_{n_s}-E_\pi}
\left(1-e^{-(E_{n_s}-E_\pi)T_b}\right)\,.\label{eq:XE+}
\end{eqnarray}
We use the time dependence to subtract the exponential terms in a similar way to the corresponding subtractions for $\Delta m_K$. The spatial integral over $\vec{x}$ is implicit in eq.\,(\ref{eq:XE}).

\begin{figure}[t]
\begin{center}\vspace{-0.05in}
\includegraphics[width=0.4\hsize]{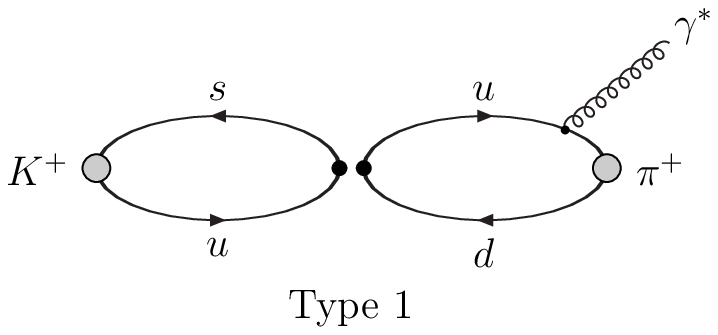}\quad\quad\includegraphics[width=0.40\hsize]{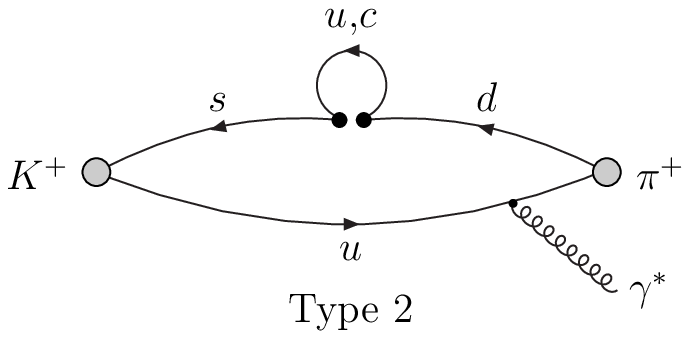}\\[0.1in] 
\includegraphics[width=0.4\hsize]{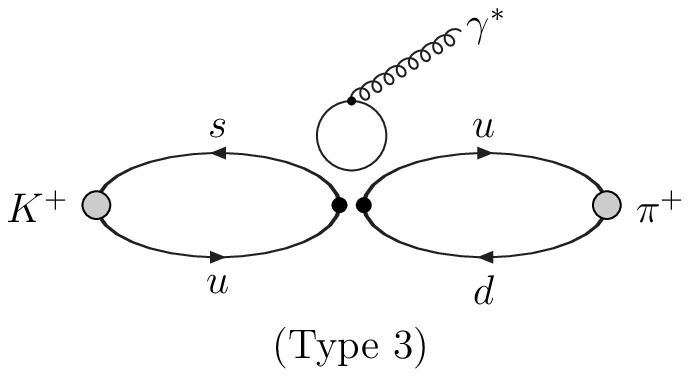}\quad\includegraphics[width=0.40\hsize]{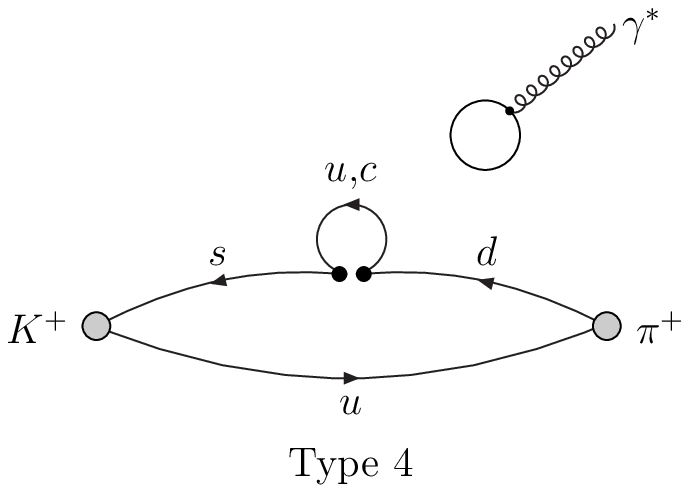}\\[0.1in]\includegraphics[width=0.4\hsize]{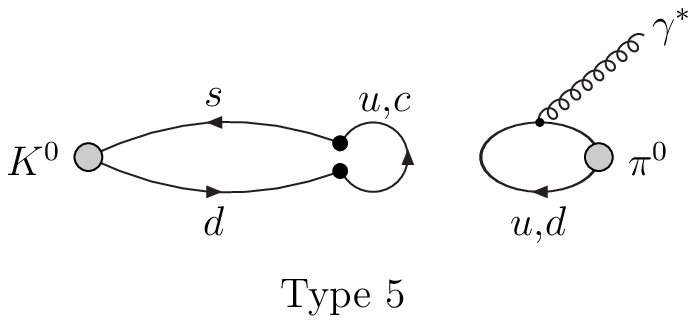}
\end{center}
\caption{Sample diagrams which need to be evaluated to determine the amplitudes for $K\to\pi\,\ell^+\ell^-$ decays. For diagrams of type 1, 2 and 5 the photon can be emitted of any internal quark line. Diagrams of type 1 - 4 contribute to both $K^+\to\pi^+\ell^+\ell^-$ and $K^0\to\pi^0\ell^+\ell^-$ decays. The diagrams of type 5 only contribute to $K^0$ decays.\label{fig:rarekaondiags}}
\end{figure}

The weak $H_W$ for this calculation is given by 
\begin{equation}\label{eq:HWrare}
H_W=\frac{G_F}{\sqrt{2}}\,V_{us}^\ast V_{ud}\,\left[C_1(Q_1^u-Q_1^c)+C_2(Q_2^u-Q_2^c)\right]\,,
\end{equation}
where 
\begin{equation}\label{eq:Q12}
Q_1^q=(\bar{s}_i\gamma^\mu(1-\gamma_5)d_i)(\bar{q}_j\gamma_\mu(1-\gamma_5)d_j)\quad\textrm{and}\quad
Q_2^q=(\bar{s}_i\gamma^\mu(1-\gamma_5)d_j)(\bar{q}_j\gamma_\mu(1-\gamma_5)d_i)\,,
\end{equation}
and $(i,j)$ are colour labels. Sample diagrams which have to be evaluated to determine the amplitudes for $K\to\pi\,\ell^+\ell^-$ decays are presented in Fig.\,\ref{fig:rarekaondiags}. 

\begin{figure}[t]
\begin{center}
\includegraphics[width=0.30\hsize]{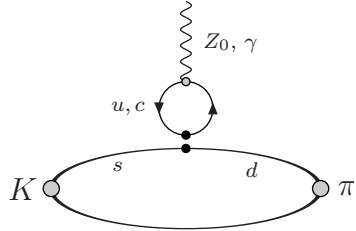}
\caption{Contribution in spite of power counting is logarithmically divergent in the ultraviolet, in spite of na\"ive dimensional counting. $H_W$ is represented by the two small filled circles.\label{fig:shortdistance}}
\end{center}
\end{figure}

The authors of ref.\,\cite{Isidori:2005tv} investigated the ultraviolet behaviour as the current $J_\mu$ approaches $H_W$. 
For illustration consider the diagram of type 2 shown in Fig.\,\ref{fig:shortdistance}, redrawn using the Fierz identity.
Dimensional counting allows for a quadratic divergence in such diagrams but conservation of the vector current suggests that the degree of divergence is reduced by 2 to result in a logarithmic divergence. For this to be the case the conserved lattice vector current must be used in the simulations. This was checked in an explicit one-loop perturbative calculation for Wilson and Clover fermion actions in~\cite{Isidori:2005tv}. This absence of power divergences does not require the use of the GIM mechanism and for a chiral symmetric formulation of lattice QCD, such as DWF, the same applies for the axial current. If the calculations are performed in the four-flavour theory, i.e. with charm quarks, then the GIM mechanism also cancels the logarithmic divergence present in this diagram.

\begin{figure}
\begin{center}
\includegraphics[width=0.35\hsize]{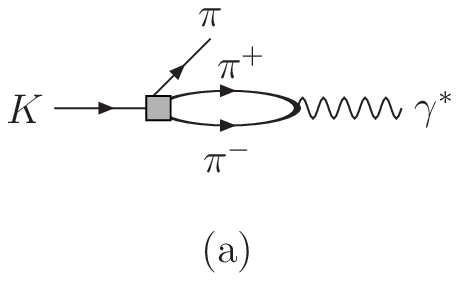}\qquad\qquad\includegraphics[width=0.45\hsize]{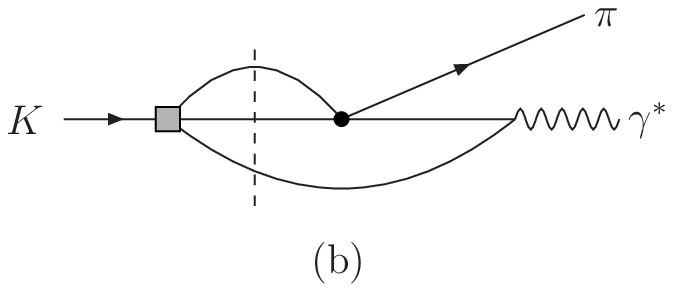}
\end{center}
\caption{Two chiral perturbation theory diagrams contributing to the decay $K\to\pi\gamma^\ast$. For $K_S$ decays there is an additional topology with a purely gluonic intermediate state. \label{fig:chpt}}
\end{figure}

In chiral perturbation theory the leading contribution to the amplitude comes from diagram (a) in fig.\,\ref{fig:chpt}. For $q^2<4m_\pi^2$, the two pions are below threshold and there are no finite-volume corrections which decrease as powers of the volume, leaving only ones which fall exponentially. Indeed there are no power corrections coming from on-shell two-pion intermediate states. Inserting the decomposition $ \bra{\pi(p_1)}V_{\mu}\ket{\pi(p_2)\pi(p_3)}=\epsilon_{\mu\nu\rho\sigma}
        p_1^{\nu}p_2^{\rho}p_3^{\sigma}F(s,t,u)\,,$
    where $s=(p_1+p_2)^2$, $t=(p_1-p_3)^2$ and $u=(p_2-p_3)^2$, into the correlation function and integrating over the phase-space of the two-pion state leads us to contract the $\epsilon$ tensor with two independent momenta ($p_K$ and $p_\pi$) and the Lorentz index of the $\gamma^\ast$ and hence we get zero. This leaves us with possible power corrections from 3-pion intermediate states (see for example the diagram in Fig.\,\ref{fig:chpt}(b)). Such diagrams are higher order in chiral perturbation theory suggesting that they are relatively small, but this requires further investigation. 
 
\begin{figure}[t]
\begin{center}
\includegraphics[width=0.35\hsize]{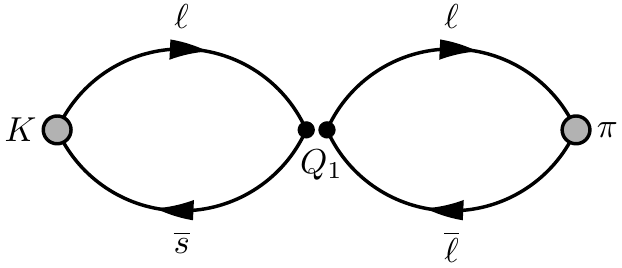}\qquad
\includegraphics[width=0.35\hsize]{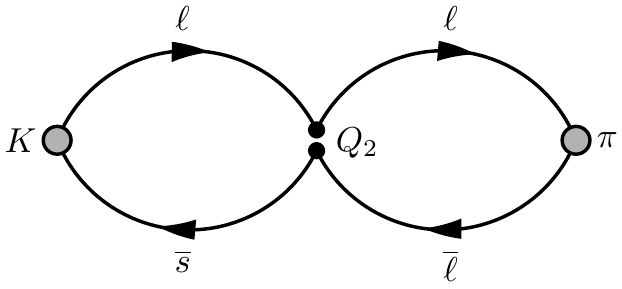}\end{center} \vspace{-0.2in}\mbox{}\hspace{1.8in} {\large${W}$}\hspace{2.2in}{\large${C}$}
\begin{center}
\includegraphics[width=0.35\hsize]{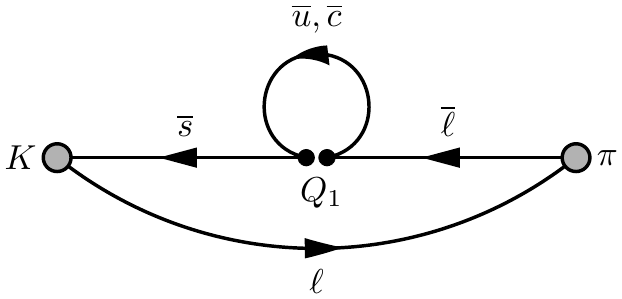}\quad
\includegraphics[width=0.35\hsize]{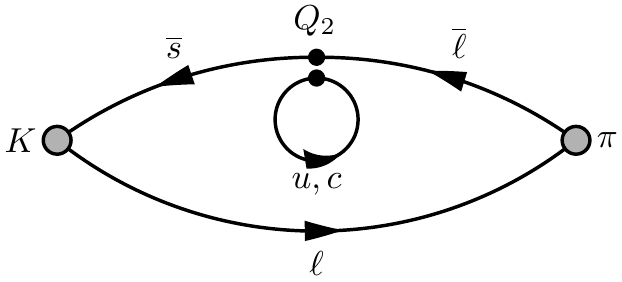}\end{center} \vspace{-0.1in}\mbox{}\hspace{1.8in}{\large${S}$}\hspace{2.2in}{\large${E}$}
\caption{The four three-point diagrams contributing to the $K^+\to\pi^+$ transition ($W$=wing, $C$=connected, $S$=saucer and $E$=eye).  \label{fig:rarediags}}
\end{figure}

\begin{figure}[t]
\begin{center}
\includegraphics[width=0.485\hsize]{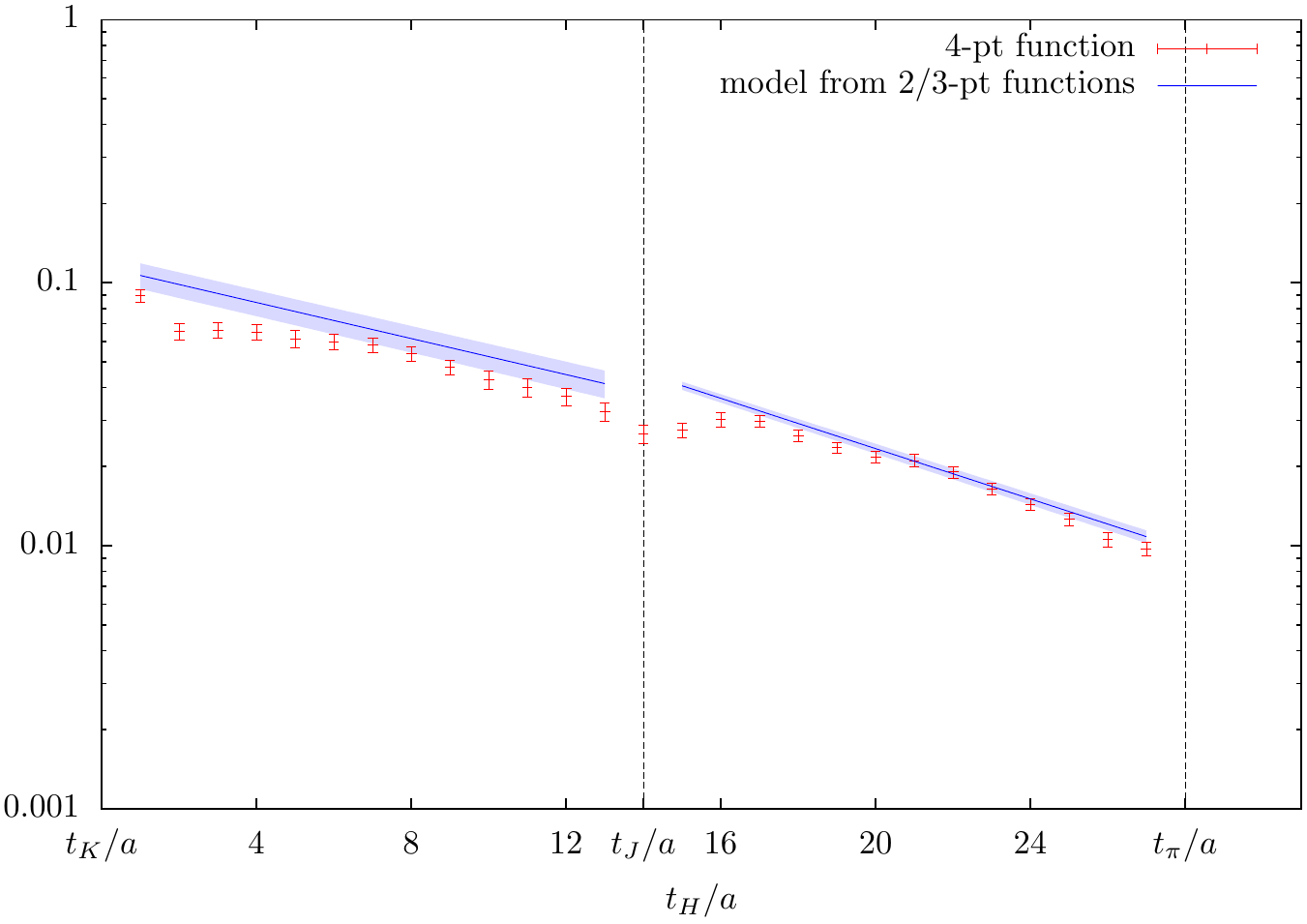}
\includegraphics[width=0.485\hsize]{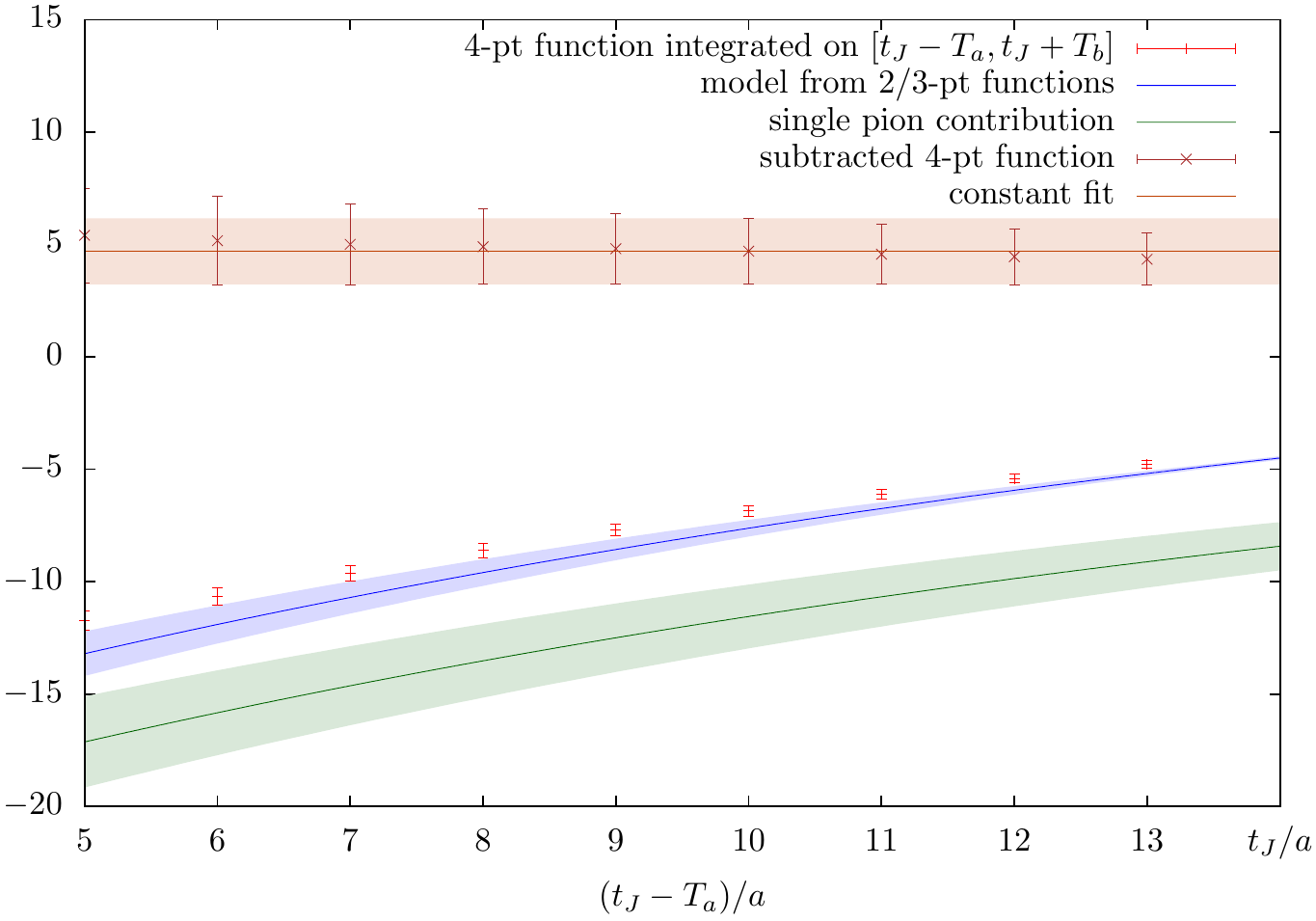}
\end{center}
\vspace{-0.3in}\caption{The left-hand plot is the unintegrated correlation function with the kaon and pion sources at $t=0$ and 28 respectively and the vector current fixed at $t=14$. The $x$-coordinate is the position of $H_W$. The right-hand plot represent the correlation function integrated from the $x$-coordinate to 23. The green band represents the exponentially growing term in  (\protect\ref{eq:XE-}) and the horizontal orange band has this removed.\label{fig:rareknum}}
\end{figure}

There are very many diagrams to evaluate. For example for $K^+$ decays we need to evaluate the graphs obtained by inserting the electromagnetic current at all possible locations in the three-point diagrams shown in fig.\,\ref{fig:rarediags} (and adding the disconnected diagrams). For the first exploratory numerical study we have only considered the $W$ and $C$ diagrams. The numerical study is being performed on $24^3\times 64$ RBC-UKQCD ensembles using 2+1 flavours of Domain Wall Fermions and the Iwasaki gauge action with $m_\pi\simeq 420$\,MeV and $a^{-1}\simeq 1.73$\,fm.\cite{Allton:2008pn}. The conserved, 5-dimensional, vector current is used. We see from eq.\,(\ref{eq:Ti}) that the matrix element vanishes when the kaon and pion are both at rest and so we have to give one or both of the mesons a momentum. In fig.\,\ref{fig:rareknum} I present some preliminary results with the choice $\vec{k}=(1,0,0)\,2\pi/L$ and $\vec{p}=\vec{0}$. In both plots the kaon and pion interpolating operators are at $t_k=0$ and $t_\pi=28$  and the vector current is placed at $t_J=14$. The left-hand plot is the unintegrated correlation function and the $x$-axis is $t_H$, the position of the weak Hamiltonian. The blue band is the result obtained assuming that only the ground-state intermediate state contributes to $X_{E_-}$ and $X_{E_+}$ and we see that the correlation function is approximately saturated by the ground-state contributions when $t_H$ is away from the remaining operators. In the right-hand plot of fig.\,\ref{fig:rareknum} we present the integrated correlator (\ref{eq:XE}) with $T_b=9$ so that the integral over $t_H$ is from the $x$-coordinate to 23. It appears that the subtraction of the exponentially growing term (in this case arising from the pion intermediate state) can be performed and a result obtained. This project is still in its early stage, but these initial results are very encouraging.

\section{Electromagnetic Corrections to Weak Matrix Elements}\label{sec:fpiem}

\begin{table}[t]
\begin{center}
\begin{tabular}{cccccc}
$f_\pi$&$f_K$&$f_D$&$f_{D_s}$&$f_B$&$f_{B_s}$\\ 
130.2(1.4)&156.3(0.8)&209.2(3.3)&248.6(2.7)&190.5(4.2)&227.7(4.5)
\end{tabular}\end{center}
\vspace{-0.15in}\caption{Results for the leptonic pseudoscalar decay constants~\cite{Aoki:2013ldr}.The results are presented in MeV.\label{tab:decayconstants}}
\end{table}

At this conference, Antonin Portelli has reviewed recent calculations of the hadronic spectrum in which electromagnetic effects are included. Here I present a proposed procedure to include electromagnetic effects in weak matrix elements~\cite{Carrasco:2015xwa}. The motivation for this is that some of the lattice results for these matrix elements are now being quoted with
$O(1\%)$ precision,  e.g. in tab.\,\ref{tab:decayconstants} I list the values of the decay constants compiled by the FLAG collaboration~\cite{Aoki:2013ldr}. We therefore need to start considering electromagnetic (and other isospin breaking) effects if we are to use these results to extract CKM matrix elements at a similar precision. The principal new feature when calculating electromagnetic effects in decay (and scattering) processes is the presence of infrared divergences, and it is the treatment of infrared divergences in lattice simulations which is the main subject of ref.\,\cite{Carrasco:2015xwa}.

In the following discussion, for illustration I consider leptonic decays of the pion but the discussion is general and can be easily generalised to other leptonic and semileptonic decays. We do not rely on chiral perturbation theory (ChPT), but for  a ChPT based discussion of $f_\pi$, see \cite{Gasser:2010wz}.

\subsection{Infrared divergences}\label{subsec:irdivs}

The presence and cancellation of infrared divergences in physical processes in QED has been understood for over 75 years now\,\cite{Bloch:1937pw}. Using the leptonic decays of the pion for illustration, at $O(\alpha)$ there is an infrared divergence in the amplitude for the process $\pi^+\to\ell^+\nu_\ell$ ($\ell=e,\mu$) which arises from the integral over the momentum of the virtual photon and in the rate for the process $\pi^+\to\ell^+\nu_\ell\gamma$ from the region of phase-space in which the real final-state photon is soft. As is well known\,\cite{Bloch:1937pw}, the infrared divergences cancel between contributions to the rate from diagrams with real and virtual photons. For the foreseeable future it will be sufficient to restrict the calculations to $O(\alpha)$ and only consider diagrams with a single virtual or real photon.

From the above, we see that when calculating $O(\alpha)$ corrections to the leptonic decays it is necessary to consider together the processes $\pi^+\to\ell^+\nu_\ell$ and $\pi^+\to\ell^+\nu_\ell\gamma$ and to calculate the combined width
\begin{equation}
\Gamma(\pi^+\to\ell^+\nu_{\ell}(\gamma))=\Gamma(\pi^+\to\ell^+\nu_{\ell})+\Gamma(\pi^+\to\ell^+\nu_{\ell}\gamma)
\equiv\Gamma_0+\Gamma_1\,,
\end{equation}
where the subscript ${\scriptsize 0}$ or ${\scriptsize 1}$ denotes the number of photons in the final state. The question for our community is how best to combine this understanding with lattice calculations of non-perturbative hadronic effects and it is this question which we now begin to tackle.
I repeat that this is a generic problem which needs to be solved if electromagnetic corrections are to be included in the evaluation of decay processes.

At $O(\alpha^0)$, i.e. without electromagnetic corrections, there is no photon in the final state and the total width is given by
\begin{equation}
\Gamma(\pi^+\to\ell^+\nu_{\ell})=\frac{G_F^2\,|V_{ud}|^2f_\pi^2}{8\pi}\,m_\pi\,m_\ell^2\left(1-\frac{m_\ell^2}{m_\pi^2}\right)^{\!\!\!2}\,.
\end{equation}
In this case the hadronic effects can be parametrised by a single number, the leptonic decay constant $f_\pi$ and  lattice calculations of decay constants have been performed for several decades (see tab.\,\ref{tab:decayconstants}). Once the decay constants have been determined in lattice simulations they can be combined with the experimental measurements of the widths to obtain the corresponding CKM matrix elements. Such a parametrisation of the widths in terms of a single decay constant does not apply at $O(\alpha)$.

In principle, particularly as techniques and resources improve in the future, it may become possible (and perhaps better) to compute $\Gamma_1$ over a large range of photon energies using lattice simulations. At this stage however, we do not propose to compute $\Gamma_1$ nonperturbatively. Instead we consider only real photons which are sufficiently soft for the point-like (pt) approximation to be valid. A cut-off $\Delta E$ of $O(10\,$-$\,20\,\textrm{MeV})$ appears to be appropriate both theoretically and experimentally\,\cite{Ambrosino:2005fw,Ambrosino:2009aa}. At $O(\alpha)$ we therefore propose to calculate $\Gamma_0+\Gamma_1(\Delta E)$ for sufficiently small $\Delta E$ that $\Gamma_1(\Delta E)$ can be calculated using perturbation theory, neglecting the structure dependence of the pion and treating it as an elementary pseudoscalar meson.

In order to facilitate control of the cancellation of infrared divergences we write: 
\begin{equation}\label{eq:master}
\Gamma_0+\Gamma_1(\Delta E)=\lim_{V\to\infty}(\Gamma_0-\Gamma_0^{\mathrm{pt}})+
\lim_{V\to\infty}(\Gamma_0^{\mathrm{pt}}+\Gamma_1(\Delta E))\,.
\end{equation}
The $V\to\infty$ limits are included as a reminder that lattice calculations are performed in finite volumes and an extrapolation to infinite volume is then taken. $\Gamma_0^\mathrm{pt}$ is an unphysical quantity. It is the width obtained by treating the pion as as elementary pseudoscalar which can therefore be calculated in perturbation theory.

By taking $\Delta E$ to be sufficiently small that structure dependent terms can be neglected, the second term on the right-hand side of eq.\,(\ref{eq:master}) can be calculated in perturbation theory directly in infinite volume. It is infrared convergent by the Bloch-Nordsieck mechanism~\cite{Bloch:1937pw}, but it does contain terms proportional to $\log\Delta E$.

$\Gamma_0$ is calculated nonperturbatively  whilst $\Gamma_0^\mathrm{pt}$ is calculated in perturbation theory, both in finite volumes. In the infrared region $\Gamma_0\to\Gamma_0^\mathrm{pt}$ so that the first term on the right-hand side of eq.\,(\ref{eq:master}) is also free of infrared divergences.
The subtraction in the first term is performed for each of the discrete photon momenta $k$ and then the sum over $k$ is performed. Note that the contribution of the zero mode $k=0$ naturally cancels in this subtraction.

\subsection{$\mathbf{G_F}$ at $\mathbf O(\alpha)$}\label{subsec:GF}

The predicted widths depend on the Fermi constant $G_F$ which is practice is inferred from the measured value of the muon lifetime, $\tau_\mu$. At $O(\alpha)$ one has to take care that the procedure used to determine $G_F$ is consistent with that used in making predictions for the decays of hadrons. 
Conventionally $G_F$ is determined from the expression~\cite{Berman:1958ti,Kinoshita:1958ru}
\begin{equation}\label{eq:mulifetime}
\frac{1}{\tau_\mu}=\frac{G_F^2m_\mu^5}{192\pi^3}\left[1-\frac{8m_e^2}{m_\mu^2}\right]\left[1+\frac{\alpha}{2\pi}\left(\frac{25}{4}-\pi^2\right)\right].
\end{equation}This expression can be viewed as the definition of $G_F=1.16632(2)\times 10^{-5}\,\mathrm{GeV}^{-2}$. 

\begin{figure}[t]
\begin{center}
\includegraphics[width=0.25\hsize]{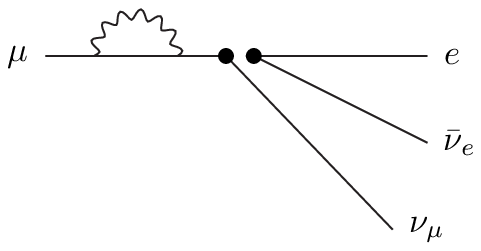}\qquad
\includegraphics[width=0.25\hsize]{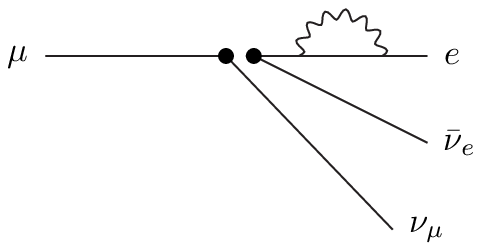}\qquad
\includegraphics[width=0.25\hsize]{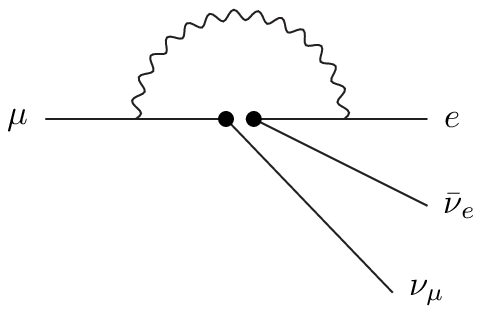}
\end{center}
\caption{The three diagrams with a virtual photon contributing at $O(\alpha)$ to the muon lifetime in eq.\,(\protect\ref{eq:mulifetime}).\label{fig:mulifetime}}
\end{figure}

In writing $\tau_\mu$ as in eq.\,(\ref{eq:mulifetime}) many EW corrections have been absorbed into the definition of $G_F$; the explicit $O(\alpha)$ corrections come from the three diagrams in the effective theory shown in fig.\,\ref{fig:mulifetime} together with the diagrams with a real photon.
The diagrams in fig.\,\ref{fig:mulifetime} are evaluated in the $W$-regularisation~\cite{Sirlin:1980nh} in which the Feynman-gauge photon propagator is modified by 
\begin{equation}
\frac{1}{k^2}\to\frac{M_W^2}{M_W^2-k^2}\,\frac{1}{k^2}\,.\end{equation}
Note that 
\begin{equation}
\frac{1}{k^2}=\frac{1}{k^2-M_W^2}+\frac{M_W^2}{M_W^2-k^2}\,\frac{1}{k^2}\,,
\end{equation}
and the contributions from the first term, which are generally ultraviolet divergent, are absorbed in the definition of $G_F$, whereas those from the second term in the three diagrams in fig.\,\ref{fig:mulifetime} result in the explicit $O(\alpha)$ term in (\ref{eq:mulifetime}).

\subsection{Proposed calculation of $\boldmath {\Gamma_0-\Gamma_0^{\mathrm{pt}}}$}

Most (but not all) of the EW corrections which are absorbed in $G_F$ are common to other processes (including pion decay). This leads to a factor in the amplitude of $(1+3\alpha/4\pi(1+2\bar{Q})\log M_Z/M_W)$, where $\bar{Q}=\frac12(Q_u+Q_d)=1/6$~\cite{Sirlin:1981ie,Braaten:1990ef}. This is a tiny correction, but one which can be included. We therefore need to calculate the pion-decay diagrams in the effective theory with 
\begin{equation}\label{eq:Heff}
H_\mathrm{eff}=\frac{G_F}{\sqrt{2}}\,V_{ud}^\ast\left(1+\frac{\alpha}{\pi}\log\frac{M_Z}{M_W}\right) (\bar{d}_L\gamma^\mu u_L) (\bar{\nu}_{\ell,\,L}\,\gamma_\mu \ell_L)\,\end{equation}
in the $W$-regularization.

Of course in practical calculations the lattice spacing $a\gg 1/M_W$ so that we cannot perform the simulations directly in the W-regularization. However the operators in the $W$ and bare lattice and regularisations can be matched using perturbation theory.
Thus for example, with the Wilson action for both the gluons and fermions:
\begin{eqnarray}
O_1^\mathrm{W-reg}&=&\left(1+\frac{\alpha}{4\pi}\left(2\log a^2M_W^2-15.539\right)\right)O_1^\mathrm{bare} + \frac{\alpha}{4\pi}\ \left(
0.536\,O_2^\mathrm{bare} \right.\nonumber \\  
&&  \left.   \hspace{0.3in}+1.607\,O_3^\mathrm{bare}-3.214\,O_4^\mathrm{bare}-0.804\,O_5^\mathrm{bare} \right) \,,\label{eq:O1Wreg}
\end{eqnarray}
where 
\begin{align}
O_1&= (\bar{d}\gamma^\mu (1-\gamma^5)u)\,(\bar{\nu}_\ell\gamma_\mu(1-\gamma^5)\ell)&O_2&=
(\bar{d}\gamma^\mu (1+\gamma^5)u)\,(\bar{\nu}_\ell\gamma_\mu(1-\gamma^5)\ell)\nonumber\\ 
O_3&= (\bar{d}(1-\gamma^5)u)\,(\bar{\nu}_\ell(1+\gamma^5)\ell)&O_4&= (\bar{d}(1+\gamma^5)u)\,(\bar{\nu}_\ell(1+\gamma^5)\ell)\nonumber\\ 
O_5&=(\bar{d}\sigma^{\mu\nu}(1+\gamma^5) u)\,(\bar{\nu}_\ell\sigma_{\mu\nu}(1+\gamma^5)\ell)\,.\label{eq:O1to5}
\end{align}
The presence of the 5 operators in (\ref{eq:O1to5}) is a manifestation of the breaking of chiral symmetry in the Wilson theory.

\begin{figure}[t]
\begin{center}
\includegraphics[width=0.27\hsize]{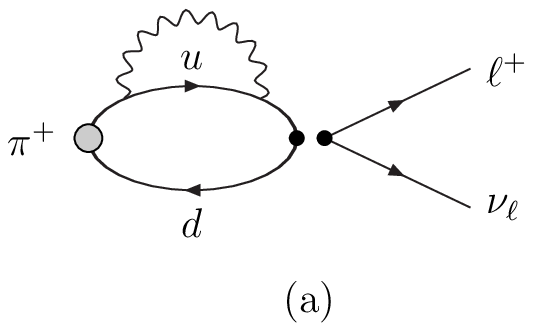}\quad
\includegraphics[width=0.27\hsize]{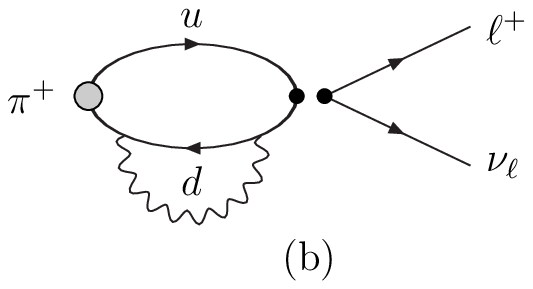}\quad
\includegraphics[width=0.27\hsize]{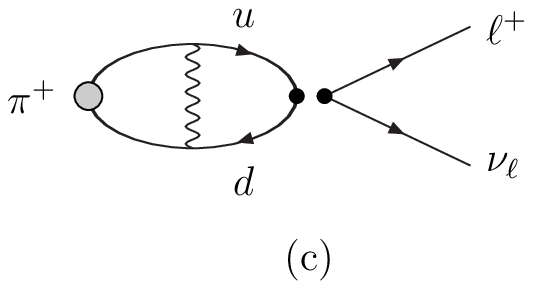}\\[0.1in]
\includegraphics[width=0.27\hsize]{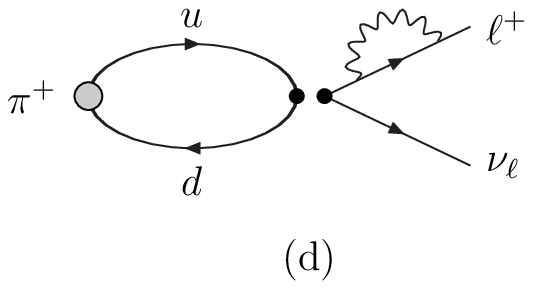}\quad
\includegraphics[width=0.27\hsize]{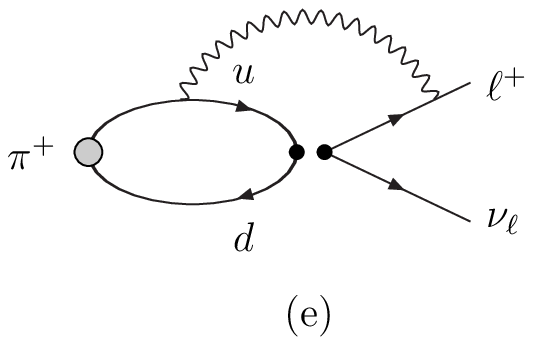}\quad
\includegraphics[width=0.27\hsize]{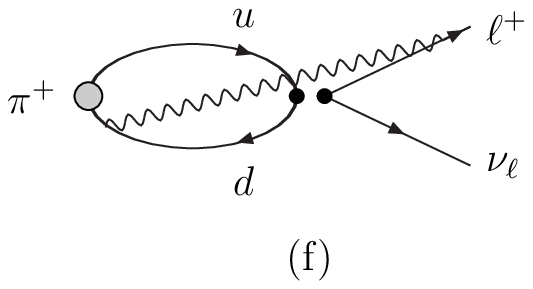}
\end{center}
\caption{Connected diagrams contributing to the correlation function at $O(\alpha)$ from which $\Gamma_0$ is determined. The shaded circle represents the interpolating operator for the pion and the two filled circles represent one of the four-fermion operators in (\protect\ref{eq:O1to5}).  The wiggly line represents the photon. \label{fig:connected}}
\end{figure}

The procedure is therefore to calculate the matrix elements of the bare lattice operators and use perturbative relations such as (\ref{eq:O1Wreg}) to match them to the W-regularization and eq.\,(\ref{eq:Heff}) is the corresponding effective Hamiltonian. The details of the calculation of the correlation function from which $\Gamma_0-\Gamma_0^\mathrm{pt}$ is determined can be found in~\cite{Carrasco:2015xwa}. The connected diagrams  can be found in Fig.\,\ref{fig:connected} and there are also disconnected diagrams to be evaluated. A relatively minor, but welcome nonetheless, simplification is that the leptonic wave function renormalisation cancels in the subtraction $\Gamma_0-\Gamma_0^\textrm{pt}$ and so does not have to be evaluated explicitly. The presence of diagrams such (e) and (f), in which the photon links the hadronic and leptonic components, confirms that the result cannot simply be written in terms of a generalised decay constant. In~\cite{Carrasco:2015xwa} it is demonstrated that the necessary Euclidean-Minkowski continuation can indeed be performed and it is shown how the matrix elements can be extracted.

\subsection{Calculation of $\Gamma^{\textrm{pt}}=\Gamma_0^{\textrm{pt}}+\Gamma_1^{\textrm{pt}}$}

The calculation of $\Gamma^{\textrm{pt}}=\Gamma_0^{\textrm{pt}}+\Gamma_1^{\textrm{pt}}$ is completely perturbative.
The total width, $\Gamma^{\textrm{pt}}$ was calculated in 1958/9 using a Pauli-Villars regulator for the UV divergences and $m_\gamma$ for the infrared divergences~\cite{Berman:1958ti,Kinoshita:1959ha}. This is a very useful check on our perturbative calculation. We add the label ${\scriptsize\mathrm{pt}}$ on $\Gamma_1$ here because the integrations include photon momenta for which the structure dependent effects for a real pion should be included, whereas in these calculations the pion is treated as an elementary point-like particle. For the proposed evaluation of the decay width, in which $\Delta E$ is sufficiently small that structure dependent contributions can be neglected, $\Gamma_1(\Delta E)\simeq\Gamma_1^\textrm{pt}(\Delta E)$. However, in order to cross-check our perturbative calculation with earlier results we keep $\Delta E$ unconstrained here.

In ref.\,\cite{Carrasco:2015xwa} we calculate $\Gamma_0^{\textrm{pt}}+\Gamma_1^{\textrm{pt}}(\Delta E)$ for a general value of $\Delta E$ using the 
following Lagrangian for the interaction of a point-like pion with the leptons:
\begin{eqnarray*}
\mathcal{L}_{\pi\textrm{-}\ell\textrm{-}\nu_\ell} &=&i\, G_F f_\pi  V_{ud}^* ~  \left\{(\partial_\mu -ieA_\mu) \pi\right\}  \, \left\{\bar{\psi}_ {\nu_\ell} \frac{1+\gamma_5}{2} \gamma^\mu \psi_\ell\right\}  + 
\mathrm{h.c.}\,.   
\end{eqnarray*}
with the corresponding Feynman rules:
\begin{eqnarray}
\includegraphics[width=0.5\hsize]{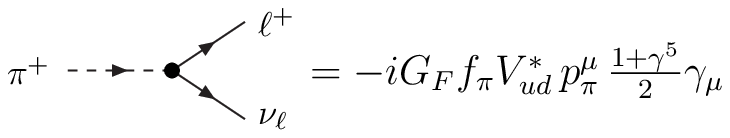}\\
\includegraphics[width=0.5\hsize]{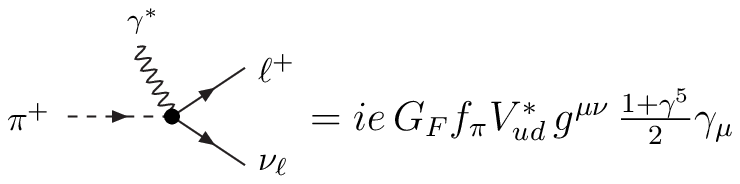}\,\cdot\nonumber
\end{eqnarray}
We believe that this is a new result, and by setting $\Delta E$ to its maximum value, $\Delta E=m_\pi/2\times(1-m_\ell^2/m_\pi^2)$ we recover the earlier result as a check. Thus the calculation of the second term on the right-hand side of eq.\,(\ref{eq:master}) is complete.

\subsection{Concluding remarks}\label{subset:summary}
In this section we have seen how the cancellation of infrared diveregences \`a la Bloch-Nordsieck can be implemented in lattice calculations of the electromagnetic corrections to decay widths. Although challenging, the method is within reach of present simulations and we now plan to implement the procedure in an actual numerical computation. 

Remaining theoretical issues to be investigated include the expected power-like finite-volume corrections in $\Gamma_0-\Gamma_0^\textrm{pt}$. Since infrared divergences cancel in this difference and in a finite volume so does the contribution of the zero momentum mode $k=0$, where $k$ is the momentum of the photon, we would expect the finite-volume effects to be similar to those in the spectrum. 

The matching factors between the bare lattice operators and those in the W-regularization have been calculated for the Wilson and some related lattice actions in~\cite{Carrasco:2015xwa} at $O(\alpha)$. As the numerical computations become performed it would be interesting to calculate the terms of $O(\alpha\,\alpha_s(a))$.

In order to estimate the size of the structure dependent effects in $\Gamma_1(\Delta E)$ without computing them in a lattice simulation some model input or approximations are necessary. In ref.\,\cite{Carrasco:2015xwa} we use chiral perturbation theory to estimate these effects for the pion and kaon. For decays into a muon the effects are negligible even for $\Delta E$ much larger than our notional 20\,MeV. For the electron, because of a kinematic enhancement proportional to $1/m_\ell^2$, the effects start to become visible at around 20\,MeV. For heavy mesons, and in particular for the $B$-meson, there is a natural small energy scale, the hyperfine splitting $m_{B^\ast}-m_B\simeq 45\,\textrm{MeV}$. Thus a $B$-meson can emit a relatively soft photon and the hadronic matter can rearrange itself into the vector $B^\ast$ meson. We would expect this to result in an earlier onset of significant structure dependent corrections and this needs to be investigated further.

In the future one can envisage relaxing the condition $\Delta E\ll \Lambda_{\mathrm{QCD}}$ and to include the emission of real photons with energies which do resolve the structure of the initial hadron. Such calculations can be performed in Euclidean space under the same conditions as above, i.e. providing that there is a mass gap. The natural generalisation of the present strategy would be to replace~(\ref{eq:master}) by
\begin{equation}\label{eq:master2}
\Gamma_0+\Gamma_1(\Delta E)=\lim_{V\to\infty}(\Gamma_0-\Gamma_0^{\mathrm{pt}})+\lim_{V\to\infty}(\Gamma_1(\Delta E)-\Gamma_1^{\mathrm{pt}}(\Delta E))+
\lim_{V\to\infty}(\Gamma_0^{\mathrm{pt}}+\Gamma_1^\textrm{pt}(\Delta E))\,.
\end{equation}
Whereas in (\ref{eq:master}) we had envisaged $\Delta E$ being sufficiently small so that $\Gamma_1(\Delta E)$ can be calculated in the point like approximation this is no longer the case here. Note that each of the three terms in (\ref{eq:master2}) is separately infrared finite and also that the results from our perturbative 
calculation of $\Gamma_0^{\textrm{pt}}+\Gamma_1^{\textrm{pt}}(\Delta E)$ will still be necessary. 

\section{Summary}\label{sec:concs} 
Standard lattice calculations of nonperturbative QCD effects in weak decays or hadronic structure are based on the evaluation of matrix elements of local composite operators. In this talk I have discussed the evaluation of (nonlocal) long-distance effects through the computation of time ordered products of two local operators integrated over their positions. The general framework has been applied to (i) the evaluation of $\Delta m_K=m_{K_L}-m_{K_S}$  (for prospects for evaluating the indirect CP-violation parameter $\epsilon_K$ see~\cite{Christ:2014qwa}); (ii) the computation of the amplitudes of the rare kaon decays $K\to\pi\ell^+\ell^-$ (for the evaluation  of the $K\to\pi\nu\bar{\nu}$ decay amplitudes see~\cite{xu}) and (iii) the calculation of the $O(\alpha)$ electromagnetic effects in weak decays of pseudoscalar mesons. The novelty in the last case is the presence of infrared divergences which cancel between contributions to the width with real and virtual photons.

For $\Delta m_K$ the early results are very promising and strongly suggest that it will soon be possible to perform a calculation at physical kinematics. For the rare kaon decays the numerical studies are at an earlier stage but are progressing well and for the electromagnetic corrections they are just beginning. These are very exciting times.

As the last speaker at this conference it is my privilege and pleasure, on behalf of all the participants, to thank Norman Christ, Bob Mawhinney, Peter Petreczky and all the members of the local organising committee for creating such a stimulating, enjoyable and beautifully organised meeting. 

\subsection*{Acknowledgements} I warmly thank my colleagues from the RBC \& UKQCD collaborations and my collaborators from ref.~\cite{Carrasco:2015xwa} for their help in developing my understanding of the material presented in this talk. This work was partly supported by UK STFC Grants ST/G000557/1 and ST/L000296/1.

\end{document}